\documentclass[11pt]{article}

\date{}

\usepackage{url}
\usepackage{cite}
\usepackage{plain}

\usepackage{wrapfig}
\usepackage{graphics}
\usepackage{picins,graphicx}
\usepackage[english]{babel}
\usepackage[leqno]{amsmath}
\usepackage{amssymb}
\usepackage{verbatim}
\usepackage[mathscr]{eucal}

\usepackage{amstext}
\usepackage{makeidx}
\usepackage{amsthm}

\usepackage{hyperref}
\hypersetup{colorlinks,%
citecolor=red,%
linkcolor=blue,%
urlcolor=black}

\makeindex

\newtheorem{theorem}{Theorem} 

\newtheorem{proposition}{Proprieta'}
\newtheorem{definition}{Definizione}
\newtheorem{notation}{Nota}
\newtheorem{ex}{Esercizio} 
\newtheorem{esempio}{Esempio}

\newcommand{\vs}{\vspace{3mm}}
 
\newcommand{\no}{\noindent} 

\newcommand{\beq}{\begin{equation}} 
\newcommand{\eeq}{\end{equation}}

\newcommand{\bex}{\begin{ex}} 
\newcommand{\eex}{\end{ex}} 

\newcommand{\bese}{\begin{esempio}} 
\newcommand{\eese}{\end{esempio}} 

\newcommand{\bpro}{\begin{proposition}} 
\newcommand{\epro}{\end{proposition}}

\newcommand{\bthe}{\begin{theorem}} 
\newcommand{\ethe}{\end{theorem}}

\newcommand{\bnote}{\begin{notation}} 
\newcommand{\enote}{\end{notation}}

\newcommand{\bdefi}{\begin{definition}} 
\newcommand{\edefi}{\end{definition}} 

\newcommand{\bc}{\begin{center}} 
\newcommand{\ec}{\end{center}}

\newcommand{\mail}[1]{\href{unina:#1}{\texttt{#1}}}

\usepackage{pdfsync}

\author{Monica De Angelis\thanks{Univ. of Naples  "Federico II", Scuola Politecnica e delle Scienze di Base.  Dip. Mat. Appl. "R.Caccioppoli", Via Cintia, Monte S. Angelo
I- 80126 Naples, Italy.
\newline\mail{modeange@unina.it}}}

\title {On diffusion  effects of the perturbed sine-Gordon equation with Neumann boundary conditions}

\begin{document}
\maketitle
\begin{abstract}
The Neumann boundary problem for the perturbed sine-Gordon equation describing  the electrodynamics of Josephson junctions has been considered. The behavior  of a viscous
term, described by a higher-order derivative with small diffusion coefficient $ \varepsilon, $ is  investigated.  The Green function  related to the linear third order operator is  determined by means of  Fourier series, and properties of rapid convergence are established. Furthermore, some classes of solutions of the hyperbolic equation have been determined, proving that there exists at least one solution whose derivatives are bounded.
Results prove that  diffusion effects 
 are bounded and tend to zero when $ \varepsilon $  tends to zero.

\vs \no {\bf{Keywords}}:{ Superconductivity; \hspace{2mm} Junctions\hspace{2mm}  Initial- boundary problems for higher order parabolic equations }

\vs \no \textbf{Mathematics Subject Classification (2000)}\hspace{1mm}44A10  \hspace{1mm}35A08,\hspace{1mm}74K30,\hspace{1mm} 35K35,\hspace{1mm}35E05

\vs \no \textbf{ PACS}{\hspace{2mm}  74.50.+r \hspace{2mm} 02.30.Jr }
\end{abstract}

\section{Introduction}
\label{intro}

 Let  us consider  the  sine-Gordon equation:

\begin{equation}  \label{11}
u_{xx}- u_{tt}  = \gamma + \sin u.
\end{equation}

This equation   models the flux dynamics in the Josephson junction where two superconductors are separated by a
thin insulating layer. Indeed, denoting by $ \lambda _L $ the London penetration depth of the superconducting electrodes, the spatial coordinate $ x$ is normalized to $ \lambda _L, $ while the time $ t $ is normalized to the inverse plasma frequency $ \omega _0 = \lambda _L /  \tilde c$    (where $\tilde c$ is  the maximum velocity of the electromagnetic waves in the junction). So,  function  $u(x,t)$ denotes the phase difference of the electrons between the top and the bottom superconductor, while  constant  $\gamma γ = j/j_0 $ (with $j_0 = $ maximum Josephson current) represents the normalized current
bias.

In this case 
superconductors are ideal, i.e. there are no quasi-particle currents and all  the electrons  form Cooper pairs.

Conversely, when a real junction is considered \cite{sco}, a  term  $\alpha  u_t$  can  denote the 
 dissipative normal electron current flow across the junction, while the flowing  of quasi-particles  parallel to  the junction can be  represented by a third term  such as  $\varepsilon \,u_{xxt}$. 
    
    So that  the following perturbed sine Gordon equation  holds: 

\begin{equation}  \label{12}
u_{xx}- u_{tt} =  \sin u + \gamma  + \alpha u_t - \varepsilon \,u_{xxt}
\end{equation} 

\noindent where the value range for $\alpha $ and
 $\varepsilon$ depends  on the material of the  real junction. Indeed, denoting by $ C, R, L_p,$ respectively, the capacitance, the resistance and the inductance  per unit length,  it results $ \alpha =1/ \omega _0 RC$   and  $ \beta = \omega _0 L_p /R.$  
 
 So,  there are cases in which  $0\leq\alpha ,\varepsilon  \leq 1$   \cite{bp,pfssu} but,
 when the resistance of the junction is so low  to completely shorten the
 capacitance, the case   $ \alpha $ large with respect
 to 1 arises \cite{ca,tin}.

Equation (\ref{12}) characterizes  rectangular or annular junctions, but other geometries can be considered such as  window Josephson junctions  (WJJ) (\cite{bc} and reference therein) or elliptic annular
Josephson tunnel junctions (EAJTJs) \cite{mm}, that  reduce to circular annular junctions as soon as    eccentricity vanishes. Moreover it is possible to consider also confocal annular Josephson tunnel junctions (CAJTJ) that are subtended by two ellipses with the same foci but do not have a constant annulus width \cite{rm}.
Besides, if an exponentially shaped Josephson junction (ESJJ)\cite{bcs2}-\cite{32} is examined, the equation achieved  is the following:  
 
\begin {equation}    \label {13} 
 \varepsilon u_{xxt}
+\, u_{xx}  -  u_{tt}  - \varepsilon\lambda u_{xt} -\lambda  u_x  -  \alpha u_t \, \, = \, \sin u - \gamma 
\end{equation}

\noindent where $ \lambda $ is a positive constant and the current due to the tapering is represented by   terms  $ \,\,\lambda \,u_x $ and $\,\lambda \, \varepsilon u_{xt} \,$. In particular,  $ \lambda u _{x} $  characterizes the  geometrical force driving the fluxons from the wide edge to the narrow edge.  

 In others cases, such as  a   semiannular or   an S-shaped  Josephson junction,  indicating by  $ L $ the length of the junction and by $ b $  an applied magnetic
field parallel to the plane of the dielectric barrier, the term $ b \cos (x \pi/\ell)$,  with $ \ell = L/ \lambda _L, $  has to be considered, too. \cite{bp,sk6}. 

Moreover, if  a harmonically
oscillating magnetic field applied parallel to the dielectric
barrier, and   a dc bias across the superconducting
electrodes are  considered,  one has \cite{sk}:

\begin{equation}    \label{14}
u_{tt}- u_{xx} + \sin u = -\gamma  - \alpha u_t   + \varepsilon \,u_{xxt}- b \sin (\omega t) \cos ( x\pi/\ell)
\end{equation}

\noindent where $\omega $ is the normalized frequency of the magnetic field
normalized to the Josephson plasma frequency $\omega _0$.
\textbf{•}

There exist numerous applications for  Josephson junctions. For example, by means of the superconducting quantum interference device (SQUID), it is possible using magnetocardiograms, to  diagnose heart and/or blood circulation problems,  while, through  magnetoencephalography -MEG- magnetic fields generated by electric currents in the brain, can be evaluated \cite {bp}.  In geophysics, on the other hand, they are used as gradiometers \cite{tin} or as gravitational wave detectors (\cite{car} and reference therein) and they play an important role in the study of the  potential virtues of  superconducting digital electronics, too \cite{f}. SQUIDs are  also used in nondestructive testing as a convenient alternative to ultra sound or x-ray methods (\cite{bp,tin}\cite{bpagano}-\cite{cla2} and reference therein).  Finally, SQUIDs can be
used as fast, switchable meta-atoms \cite{jbm}.

\subsection{Mathematical considerations}
In all the  previous   equations (\ref{12})-(\ref{14}), the following linear  operator appears:

\begin{equation}   \label{15}
{\cal L} u \equiv \,\,\partial_{xx} \,(\varepsilon
u  _{t}+u) - \partial_t( u_{t}+\alpha\,u).\,
\end{equation}

This is a third order parabolic operator that, as it is  well known, is involved  
in a vast number of realistic  mathematical models concerning  superconductivity,
neurobiology, and  viscoelasticity \cite{mda2010}-\cite{dannaf} 
where the evolution is often characterized by deep interactions
 between wave propagation and diffusion. So that, ${\cal L} $  can  be also considered  as a linear hyperbolic operator perturbed by viscous terms described by higher-order derivatives with small diffusion coefficients $ \varepsilon $.

When  $ \varepsilon  \,\equiv 0, $   the parabolic operator   turns into a  hyperbolic one:

\begin{equation}  \label{16}
{\mathcal L}_0\,U \equiv \,U_{xx}  - \partial_t(U_t+\alpha\, U),\
\end{equation}
\noindent and  the influence of the dissipative terms, represented by $\, \varepsilon \, \partial _{xxt}$, on the wave behaviour has been   estimated. 

Similar problems, for Dirichlet conditions, have already been studied in  \cite{dr1,dmr}. In particular,  when  $ \alpha=0, $ an asymptotic approximation is established by means of the  two characteristic times: slow time $ \tau = \varepsilon \,t $  and fast time $ \theta =\, t/\varepsilon $. Moreover, for  equation  (\ref{13}) in \cite{df213},  an analytical  analysis  has proved      that the  surface damping has little influence
on the behaviour of the oscillator, thus confirming numerical results already determined in \cite{bcs00}. Numerical investigations on influence of surface losses can be found in \cite{pskm},too

Here the Neumann boundary conditions added with equation (\ref{15}) are considered, and
diffusion effects have been evaluated. In order to analyze the influence of the dissipative term on the wave behavior, a rigorous estimate of operator ${\mathcal L}$
has been achieved by means of the Green function determined by Fourier series, and an evaluation
of the following difference:

\begin{equation}                \label {17}
 d(x,t,\varepsilon)\,= u(x,t,\varepsilon)  - U(x,t)  
\end{equation}

\noindent has been done. Hence, the solution of the non-linear problem related to $ d $ is determined.
Furthermore, some classes of solutions of the hyperbolic equation have been obtained, proving that there exists at least one solution whose derivatives are bounded.
Finally, since the hyperbolic equation admits solutions with limited derivatives, an
estimate for the remainder term is achieved by proving that the diffusion effects are of the order of $ \varepsilon^h $ with $h <1 $ in each interval-time $ [0, T_\varepsilon]$, with $\displaystyle T_\varepsilon = \,\min\, \{\frac{1}{N}\,\lg \bigl( \,\,  \frac{1}{\varepsilon ^{1-h}}\bigr)\,; \lg \bigl( \,\,  \frac{1}{\varepsilon ^{1-h}}\bigr)\}$ and $ N>0 $ independent from $ \varepsilon $.

 The paper is organized as follows: Section II describes the mathematical problem, and attention is fixed on  the Green function of the linear operator $ {\cal L} $ defined in (\ref{15}). In section III  properties of the Green Function $ G  $ are  pointed out  in  Theorem 1 whose   proof    can be found in appendix. Moreover, by means of  the fixed point theorem, the solution of the problem related to the remainder term    is showed. In section IV,  some explicit solutions of the non linear  hyperbolic equation  (\ref{16}) are determined.  Finally, in  section V an estimate for the remainder term  is given in  theorem 3.

\section{Statement of the problem}       

\setcounter{equation}{0}

Let $T$ be a prefixed positive value  constant and 
\begin{center} $  \Omega =\{(x,t) :  0\leq x\leq \ell 
, \  \ 0 < t \leq T \} $.\end{center}

\noindent The  Neumann  boundary  value problem for equation (\ref{12}) refers to  the phase gradient value and is  proportional to the magnetic field \cite{for,j05}. So that one has:

  \begin{equation} \label{21}
  \left \{
  \displaystyle
    \begin{array}{ll}
      \partial_{xx}(\varepsilon
u _{t}+ u) - \partial_t(u_{t}+\alpha)= \sin u+\gamma ,
 &\quad (x,t)\in \Omega, \vspace{2mm}  \\ 
    u(x,0)=h_0(x), \ \ \    u_t(x,0)=h_1(x),  &\quad   x\in [0,\ell],
  \vspace{2mm}  \\
      u_x(0,t)=\varphi_0(t), \ \ \  u_x(\ell,t)=\varphi_1(t), &\quad   0<t \leq T.
   \end{array}
     \right.
 \end{equation}

When $\varepsilon \equiv 0$, problem (\ref {21}) turns into the  Neumann   problem related   to parabolic operator ${\mathcal L}_0\,$,  which has, of course,  {\em the same initial boundary  conditions}:   

  \begin{equation}          \label{22}
  \left \{
   \begin{array}{ll}
\displaystyle
      U_{xx} -
 \partial_t(U_{t}+\alpha\,U)\,=\sin U\ +\gamma &\quad(x,t)\in \Omega,\vspace{2mm}\\ 
    U(x,0)=h_0(x), \  \  \  U_t(x,0)=h_1(x), &\quad x\in [0,\ell],\vspace{2mm}  \\
     U_x(0,t)=\varphi_0(t), \  \ \  U_x(\ell,t)=\varphi_1(t), &\quad  \ 0<t \leq T,
   \end{array}
  \right.
 \end{equation}

\noindent 
The influence of the dissipative term on the wave behavior of $\,U\,$ can be  estimated when   the difference $ d, $ defined in (\ref{17}), is  evaluated.
 
  So, let us consider  the following  problem    related to the {\em  remainder} term $ \, d:\,$

\begin{equation}          \label{23}
  \left \{
\begin{array}{ll}
     \partial_{xx} \, \,(\varepsilon
\partial_t \,+1)\,d - \partial_t(\partial+\alpha\,)d\,=\,F(x,t,d),\        & (x,t)\in \Omega,\vspace{2mm}  \\ 
    d(x,0)=0, \ \  \    d_t(x,0)=0, \  \  & x\in [0,\ell], \vspace{2mm}  \\ 
     d_x(0,t)=0, \  \ d_x(l,t)=0,  &  0<t \leq T
 \end{array}
  \right.
 \end{equation}

with  
 
 \begin{equation}                   \label{24}
\,\,{F}(x,t,d)= \sin (d+U) - \sin U - \varepsilon \,  U_{xxt}.\,
\end{equation}

The operator $ {\cal L} $  has already been 	examined  by means of convolutions of Bessel
functions and a short review can be found in \cite{uffa}, while for Dirichlet problem it has  already been studied in  \cite{mda,mda12}
  and a recent approach can be  found in \cite{cor}. Here, in order to deduce an exhaustive asymptotic analysis, it  has been   analyzed by means of  the Green function  determined by Fourier series. So, assuming

 \begin{equation}  \label{25}
 \left \{                                             
\begin{array}{ll}
  &\gamma_n=\frac{n\pi}{\ell},\qquad h_n=\frac{1}{2}(\alpha+\varepsilon\gamma_n^2),\vspace{2mm}  \\ \\ & \omega_n=\sqrt{h_n^2-\gamma_n^2}
\\  \\
& H_{n}(t)=\,\, \frac{1}{\omega_n}\,\,e^{-h_nt}\,\,
\sinh\,(\omega_nt), 
\end{array}
  \right. 
 \end{equation}

by means of standard techniques,   the Green function  $ G=G(x,t) $  of problem (\ref{23}) is given by

 \begin{equation}\label{26}                                               
G(x,t,\xi)= \frac{1}{\ell}\,\, \frac{1- e^{-\alpha \, t }}{ \,\alpha}\,\, \,\,+ \,\frac{2}{\ell•}\,\,\sum_{n=1}^{\infty}
H_{n}(t) \,  \,  \cos\gamma_n\xi\,  \, \cos\gamma_nx. 
\end{equation}

 \section{ Properties of the Green function and solution related to  the remainder term } 

 \setcounter{equation}{0}

In order to achieve the explicit solution of problem (\ref{23}),
 attention should be paid to function G. Some  properties of the  Green function have already been determined  in \cite{mda,mda12,df13} proving among other things, that function G is exponentially vanishing as $t\rightarrow\infty$. Moreover, since (\ref{26}), it results:

\begin{equation}  \label{311}
G(x,t,\xi) \leq \,  \frac{2}{\ell} \,\, \sum_{n=0}^{\infty}   \,\,H_n(t)\, \cos\gamma_n\xi\,  \, \cos\gamma_nx, 
\end{equation}

 and denoting by

\begin{eqnarray}  \label{31}
\nonumber  &\beta  \equiv  \min \,\,\biggl\{ 
\frac{1}{\varepsilon+\alpha(\ell/\pi)^2},   \ \frac{\alpha+\varepsilon
(\pi/\ell)^2}{2},   \ \alpha/2 \biggr\},\\ \\  
&\nonumber r = \frac{q-1}{q•} \,\, \,(q>1)
\end{eqnarray}

the following theorem, whose proof can be found in appendix, holds:

\textbf{Theorem 1} \label{thonG}
The function $G(x,\xi,t)$  defined in
(\ref{26}),  and all its time derivatives are continuous functions    and  there exist some positive constants  $ A_j \,\, (j \in {\sf N})$ depending on   $\alpha ,\varepsilon$ and positive constants  $ M, \, N    $    depending on $ \alpha  $  and independent from  $\,\varepsilon $    such that:

\begin{equation}  \label{32}
| \, G(x,\xi,t)|\,\leq \,\, 1/2\,\, (\,M \,\,\varepsilon^{r} +N \,\, \,e^{- \beta\,\, t}\,) 
\end{equation}

\begin{equation} \label{33}
 \left | \frac{\partial ^j G}{\partial t^j} \right |\, \leq \,A_j \,  e^{-\beta
t}, \qquad j \in {\sf N} 
\end{equation}

 Furthermore, one has:

\begin{equation}                      \label{34}
{\cal L} \,\,G \, =\partial_{xx}(\varepsilon
G _{t}+ G) - \partial_t(G_{t}+\alpha G)=0.
\end{equation}

\hbox{} \hfill \rule{1.85mm}{2.82mm}

Now, let us consider   the  nonlinear source  (\ref{24})\,$\,\displaystyle F(x,t,d)= \sin (d+U) - \sin U - \varepsilon \,  U_{xxt}\,$.  By means of standard methods related to integral equations and thanks to  the fixed point theorem, owing to basic properties of  the Green function $ G  $ and the source  $\,\, F,\,\, $  it is possible   to prove   that problem (\ref{23})-(\ref{24})  admits a unique regular solution in $ \Omega $ and it results: \cite{c}-\cite{dew13}

\begin{eqnarray}  \label{41}
 d(x,t)=\,\,  -\,\frac{1}{\ell} \int_{0}^{t} \, d\tau\,\, \int_0^\ell \,\,\bigl[\frac{1- e^{-\alpha \, (t-\tau) }}{\alpha} \bigr]\,\, F(\xi,\tau,d(\xi,\tau)) \, d\xi\ \\ \nonumber \\  \nonumber    - \,\,\frac{2}{\ell•}\,\int_0^ t d\tau\, \int_0^\ell\, H(x,\xi,t-\tau)\,
F(\xi,\tau,d(\xi,\tau))\,\,d\xi.
\end{eqnarray}

where 
\begin{equation}
H= \sum_{n=1}^{\infty}
H_{n}(t) \,  \,  \cos\gamma_n\xi\,  \, \cos\gamma_nx.
\end{equation}

\section{Explicit solutions of the hyperbolic equation }

\setcounter{equation}{0}

Let us consider the semilinear  second order equation:  

\begin{equation}                \label{51}
  U_{xx} \,- \,U_{tt} \, - \,  \alpha \,U_{t} =   \sin U \,  \ + \gamma 
\end{equation}

When $ \alpha =   \gamma  = \, 0  $,  (\ref{51}) represents  the  sine - Gordon equation  and  there is plenty of  literature   about its   classes of solution. \cite{jsb}-\cite{adc}

Now, let $\, f \,$ be  an arbitrary function, and let  us  consider  the following function $ \,\,\Pi (f)\,\,$:

\begin{equation}     \label{52}
\Pi(f) = \, 2 \, \arctan\,\, e^f \,\,  
\end{equation}

\noindent so that

\begin{equation}                \label{53}
  \sin \,  \Pi (f) \,\, = \frac{1}{\cosh (f) }, \qquad  \cos \,  \Pi (f) \,\, = -   \tanh (f) .   
\end{equation}

By means of function (\ref{52}) it is possible to find a class of solutions of  equation  (\ref{51}).

 
 Indeed, it is possible  to  verify   that the following function:

\begin{equation}
U \, = 2 \, \Pi [f(\xi)]\,  \qquad \mbox{with}\qquad \xi =\, \frac{x-t}{\alpha}
\end{equation}

 is a solution of (\ref{51}) provided that one has:

\begin{equation}  \label{55}
-\alpha\, U_{t}\, = \, \sin U \, +\, \gamma. \end{equation}

\vspace{3mm} Moreover,  since (\ref{53}) and   being $\,\, \dot \Pi = \frac{1}{\cosh f}, \,\,\,  $ 
 it results:

\[ \displaystyle -\alpha\, U_{t}\, = 2\frac{f'}{\cosh f}; \,\,\qquad \sin U \,\,=  2 \sin \Pi \cos \Pi  \,\,= \,-\,\,2 \,\frac{\tanh f }{\cosh f}. 
\]

\vspace{3mm} So, from (\ref{55}), one deduces that function $ f $ must satisfy the following equation:

\begin{equation}  \label{56}
\frac{df}{-\tanh f\, +\gamma/2\, \cosh f}\,= \,d\,\xi
\end{equation}

\vspace{3mm} When  $ 0\leq \gamma \leq 1,$  we point our attention to those cases in  which it results: 

\begin{equation}
U\,=\, 4 \arctan (\, y\,+\,\sqrt{y^2 \, +\, 1}\,).
\end{equation}

\vspace{3mm} So, let $ h $  be an arbitrary constant of integration, one obtains:

\begin{equation}
 y= \,h\, e^{-\, \xi\,}\,\,\,\,\quad\quad \qquad \qquad \mbox{when}\qquad \qquad \gamma = 0,
\end{equation}

\begin{equation}
 y= \,\frac{1-(\xi\,-h)}{1+ \xi -h}\qquad \qquad  \mbox{when}\qquad \qquad \gamma = 1.
\end{equation}

\vspace{3mm} Moreover, assuming $ \gamma^2 <1,   $ let \[ A = \, \mp \sqrt{1-\gamma^2}\,\, \qquad \mbox{and} \qquad  \delta = - 1+A.\]


\noindent  For   $ \gamma^2 \neq \delta^2, $ it results:

\begin{equation}  \label{510}
 y= \,\frac{h\, \frac{\gamma}{\delta\,\,}\,\,e^{\xi A}\,\,-\frac{\delta}{\gamma\,\,}}{1-h \, e^{\,\,\xi \,\,A}}. 
 \end{equation}


\vspace{3mm} When  $  \gamma > 1,$ it is possible to prove that 

\begin{equation}
U = 2 \,\,\arctan \biggl\{\frac{1}{\gamma}\,\, \biggl[\sqrt{\gamma^2-1}\,\, \tan \biggl(\frac{\sqrt{\gamma^2-1}}{2} \, \,\,\xi + h\,\,\biggr)\,\,-1 \,\,\biggr]\,\,\biggr\}
\end{equation}

 \vspace{3mm} Physical cases show that generally   $ \gamma \,$ is less than $ \,1,  $ and in this case, indicating by  $ \displaystyle  \eta=\sqrt{1-\gamma^2},$  it also results:

\begin{equation}                       \label {49}
U(x,t) \,\ =   \, 2\,\, \arctan \,\biggl[ \,\, \frac{\eta}{\gamma} \,\, \biggl( \frac{1+ h\,\,e^{\, \xi}}{1-  h\,\,e^{\,\, \xi}} \, - \,\, \frac{1}{\eta}\,\,\biggr) \biggr] 
\end{equation}

  So, Indicating by  
  \begin{equation}
    z=  \,\, \frac{\eta}{\gamma} \,\, \biggl( \frac{1+ h\,\,e^{\, \xi}}{1-  h\,\,e^{\,\, \xi}} \, - \,\, \frac{1}{\eta}\,\,\biggr), 
    \end{equation} 
   
one has:

\begin{equation}  \label{514}
 U_{xxt}(x,t)\,= \displaystyle{\frac{2 \,\,z_{xxt}}{ 1+z^2}\,}\,\,+ \,\frac{12 \, z  \, z_x \,z_{xx}+ 4z_x^3}{ (1+z^2)^2}\,\,\,-  \frac{ 16\, z^2 \, \,z^3_{x}}{ (1+z^2)^3}\,
 \,\,  
\end{equation}

\noindent which is bounded for all $ (x,t) \in \Omega_T, $ being

\begin{eqnarray}
\begin{split}
 &z_x = \ \frac{\eta}{\gamma \, \alpha}\,\,\biggl[\, \frac{he^\xi}{1-he^\xi}\, +\,\frac{h (1+h e^\xi)}{(1-he^\xi)^2}\biggr]; \\  \\
&z_{xx} = \frac{\eta}{\gamma \, \alpha^2}\,\,\biggl[\, \frac{he^\xi}{1-he^\xi}\, +\,\frac{h^2 e^\xi (1+ e^\xi)}{(1-he^\xi)^2} \,\,+
 \,\frac{ 2 h^2 e^\xi (1-h^2 e^{2\xi})}{(1-he^\xi)^4} \,\biggr]\\ \nonumber \\
& z_{xxt} = - \frac{\eta}{\gamma \, \alpha^3}\,\,\biggl[\, \frac{he^\xi}{1-he^\xi}\, +\,\frac{3h^2 e^{2\xi}+ h^2e^\xi}{(1-he^\xi)^2} \,\,+ \,\frac{2 h^2 e^{\xi} -6h^4 e^{3\xi}}{(1-he^\xi)^4} \, 
  + \frac{ 8 h^3 e^{2\xi} (1+h^2 e^{2\xi})}{(1-he^\xi)^5}\biggr].
\end{split}
\end{eqnarray}

\quad

\section{Estimates for the remainder term}

\setcounter{equation}{0}

Let us assume  $ \varepsilon =0  $  and let $ U  $  be  a solution of the  reduced  problem (\ref{22})

In the following we will have to refer to a known  inequality of Gronwall type due to S.M. Sardar'ly (\cite{mpf} p 359):

\vspace{3mm}\textbf{Theorem 2} 
Let $ x $ and $a_2 $ be continuous   and  $ a,\,a_1,\,\int_0^t b(t,s)ds \,$  Riemann integrable  functions on $ J=[0,\beta], $ with  $ a_1 $ and $a_2$  nonnegative on J. 

 \vspace{3mm}If

\begin{equation}
x(t) \leq a(t) + \int_0^t b(t,s)ds +  a_1(t)\int_0^t a_2(s) x(s) ds \qquad t\in J
\end{equation}

 then

\begin{eqnarray}
\nonumber  x(t)  \leq & a(t) & + \int_0^t b(t,s)ds + a_1(t)\int_0^t a(s)a_2(s) \exp\biggl( \int_s^t a_1(z) a_2(z) dz\biggr) ds \, + \\ 
\\
\nonumber & a_1(t)&\int_0^ta_2(s) \int_0^s b(t,z)dz \,\, \exp\biggl( \int_s^t a_1(z) a_2(z) dz\biggr) ds  \qquad t\in J.
\end{eqnarray}
\hbox{} \hfill \rule{1.85mm}{2.82mm}

According to this, it is possible to state:

\vspace{3mm} \textbf{Theorem 3} \label{remainder} 
Let us assume

\begin{equation}
S(t)\, = \sup_{0\leq  x \leq \ell }\, | d(x,t)|. 
\end{equation}

If there exists   a  positive constant   $ k  $ such that 

\begin{equation} \label{61}
 |U_{xxt}\,(x,t)|\, \leq \,k, 
\end{equation}

\vspace{3mm}
\noindent then there exist two  positive constants  $\Gamma $  and  $ h $  with $ \,\displaystyle \, h\,<\,\frac{q-1}{q}\, (1<q<\infty) \, $  such that, indicated   by   $ N  $   the  positive constant  defined  in (\ref{32}) and

\begin{equation}                 \label{63}
 T_ \varepsilon \,: = \,\min \biggl\{ \frac{1}{N} \lg \biggl( \,\,  \frac{1}{\varepsilon ^{1-h}}\biggr);\,\,
\lg \biggl( \,\,  \frac{1}{\varepsilon ^{1-h}}\biggr)\biggr\},
\end{equation}

  it results:

\begin{equation}   \label{64}
0\,\leq \,S(t)\,\leq\, \Gamma \,\,\varepsilon ^h
\end{equation}

 for every $ t \leq  T_ \varepsilon.  $

\vspace{3mm}{Proof}: Let us consider function  $ F $ defined in (\ref{24}):

\begin{equation}  \label{65}
\, F \,= \, \sin  \, (d+U) \, -\sin U  \,- \, \varepsilon \,U_{xxt}.   
\end{equation}

\vspace{3mm} Since  (\ref{311}) and (\ref{41}), it results:

\begin{equation} \label{66}
|d(x,t,\varepsilon)| \,\leq   \,\frac{2}{\ell•}\int_0^ t d\tau\, \int_0^l\, |G(x,t-\tau \,,\xi)| \,\,\, | F(x,\tau,\xi)|\,\,d\xi\ 
\end{equation}

\vspace{3mm} where, function $ F(x,t,u), $ according to (\ref{61}) and   (\ref{65}), satisfies the following inequalities: 

\begin{eqnarray}
&|F(x,t)| \leq \,\, |d(x,t)| +\varepsilon \,\,k \nonumber
\\ 
\\\nonumber
&|F(x,t)| \leq 2+\varepsilon \,\,k. 
\end{eqnarray}

\noindent  So that, by means of properties of the  Green function $G,  $ and in particular since $(\ref{32}), $ one obtains: 

\begin{eqnarray} \label{68}
\nonumber & |d(x,t,\varepsilon)| \,\leq  \frac{1}{\ell•}\int_0^ t d\tau \int_0^l\, M\,\varepsilon^r (2+\varepsilon k)d\xi\ + 
\\ 
\\ 
\nonumber \,\,&\ \qquad \qquad \qquad\frac{1}{\ell•}\int_0^ t d\tau\, \int_0^l\, N\, \,e^{-\beta(t-\tau)}   \,[|d(\tau,\xi)|\,\, +\varepsilon \,\,k]\,\,d\xi
\end{eqnarray}

\noindent where $ \beta\,\, \mbox{and } r $ are defined in (\ref{31}).

\vspace{3mm} Hence, it follows:

\begin{eqnarray} \label{69}
\nonumber &\,S(t)\,\leq\, \, \,(2+\varepsilon \,\,k) \,\, M \, \varepsilon^r\,\, t  \,\,+\,k\,\,\varepsilon\,\,  {N}   \int_0^ t \,  e^{\,-\,{\beta}   \,(t-z)\,}\, dz \,+
\\
\\
\nonumber &\,N\,\,  e^{\,-\,\beta  \,t\,}\,\, \int_0^ t \,  e^{\,\,{\beta}  \,z\,}\,
\,S(z)\, dz .
\end{eqnarray}

Applying theorem 2  it results:

\begin{eqnarray} \label{610}
\nonumber & S  (t)\,\,\, \leq &\,M \,\,\varepsilon^r \,({2+\varepsilon \,\,k }) \,\, t  \,\, + N \,k\,\,\frac{1-e^{-\beta \,t}}{\beta} \,\,\varepsilon \,   + 
\\ \\
\nonumber && {N}  \,  e^{-\beta t}  \int_0^t    [ N\,k\,\,\varepsilon\,\frac{1-e^{-\beta \,s}}{ \, \beta} \,\,\,  \,  + M \,\,\varepsilon^r\,\, ({2+\varepsilon \,k})\,\, 
s]  e^{\beta s} \,\, e^{N(t-s) }\,\,\, ds.
\end{eqnarray}

\noindent So, one has:

\begin{eqnarray}  \label{611}
\nonumber & S(t) \,\leq \,&\biggl[ M  \,(2+\varepsilon \,\,k ) \,\, t \, \,\, + N M  (2+\varepsilon \,k) \biggl(\frac{t}{\beta-N} +\frac{e^{-t(\beta-N)}-1}{(\beta-N)^2•}\biggr)\,\,\biggr] \varepsilon^r\,+\\
 \\
 \nonumber \\
\nonumber &&\biggl[ N \,k\,\,\frac{1-e^{-\beta \,t}}{\beta} \,\,\,  \,\, +   \frac{N^2 k \,}{\beta•}\biggl(\frac{1}{\beta-N•}+\frac{\beta\, e^{-t(\beta-N)}}{N-\beta•}\,\,  + \frac{{e^{-t\beta}}}{N•}\biggr)\,\biggr] \varepsilon 
\end{eqnarray}

\vspace{3mm}  Now,  since  for all $ t\in \Re $ one has  $t+1 \leq e^t, \, $  it results  
\[ t \,\,\varepsilon ^r \leq \varepsilon^h; \qquad \varepsilon ^r  \,e^{Nt}\leq \varepsilon^h
 \]

as soon as $ \displaystyle  h < r= \frac{q-1}{q} \,\,(q>1)\,\, \mbox{and }\,\,  t\leq T_\varepsilon. $



\hbox{} \hfill \rule{1.85mm}{2.82mm}

\textbf{ Remark 6.1} Estimate   (\ref{64}) specifies the infinite time-intervals where the effects of diffusion are of the order $\,\displaystyle \varepsilon ^h \, \,$ with $\, 0<\,h\,<1. $ Indeed, the evolution of the superconductive model  is characterized by  diffusion effects which are of the order of $ \,\displaystyle  \varepsilon ^h\,\,$ in each time-interval 
$ \, [\, 0, \, \, T_\varepsilon \,]$ with $ T_\varepsilon $ defined in (\ref{63}).

\textbf{ Remark 6.2}  Formula (\ref{514}) shows that  the class of functions    satisfying  hypotheses  of Theorem 3  
 is not empty.


%
%

%


\vspace{3mm}
\textbf{Acknowledgments} 

 This work was supported by National Group of Mathematical Physics (GNFM-INDAM)

\appendix
\section{Theorem 1 }
\setcounter{equation}{0}
Let us assume:

\begin{eqnarray}  \label{31}
\nonumber  &\beta  \equiv  \min \,\,\biggl\{ 
\frac{1}{\varepsilon+\alpha(\ell/\pi)^2},   \ \frac{\alpha+\varepsilon
(\pi/\ell)^2}{2},   \ \alpha/2 \biggr\},\\ \\  
&\nonumber r = \frac{q-1}{q•} \,\, \,(q>1)
\end{eqnarray}

The function $G(x,\xi,t)$  defined in
(\ref{26})  and all its time derivatives are continuous functions    and  there exist some positive constants  $ A_j \,\, (j \in {\sf N})$  depending on   $\alpha ,\varepsilon$ and positive constants  $ M, \, N    $    depending on $ \alpha  $  and independent from  $\,\varepsilon $    such that:

\begin{equation}  \label{A1}
| \, G(x,\xi,t)|\,\leq \,\, 1/2\,\, (\,M \,\,\varepsilon^{r} +N \,\, \,e^{- \beta\,\, t}\,) 
\end{equation}

\begin{equation} \label{A2}
 \left | \frac{\partial ^j G}{\partial t^j} \right |\, \leq \,A_j \,  e^{-\beta
t}, \qquad j \in {\sf N} 
\end{equation}

 Furthermore, one has:

\begin{equation}                      \label{A3}
{\cal L} \,\,G \, =\partial_{xx}(\varepsilon
G _{t}+ G) - \partial_t(G_{t}+\alpha G)=0.
\end{equation}
\appendix
Proof:

Since $ \alpha \,\varepsilon <1, $ indicating by $ N_1, N_2 $ the integer part of   $ \ell/(\,\pi \varepsilon ) (1\mp \sqrt{1-\alpha\varepsilon }),  $   respectively,   
 circular functions have to be distinguished from hyperbolic terms. In this case, since Taylor formula, it results $ \sqrt{1-\frac{b_n^2}{h_n^2}•}\,\,\ <1-\frac{b_n^2}{2h_n^2},$ and for all  $  n\geq 1 $ one has:

\begin{equation}
\displaystyle e^{-t(h_n-\omega_n)}\leq  e^{-h_{n}t}  \,\, e^{h_{n}\bigl(1-\frac{b_n^2}{2h_n^2•}\bigr)t} \leq  e^{\,\frac{1}{\varepsilon+\alpha(\ell/\pi)^2}} 
\end{equation}

\vspace{3mm} Moreover, indicating by  $c$  an arbitrary constant less than $ 1, $  denoting by  $ N_c   $    the integer part of $ \displaystyle \ell/(\pi \varepsilon \sqrt{c}) (1+ \sqrt{1-\alpha\varepsilon \,c}), $ for all $ \displaystyle n \geq N_c,$ it results $ \, \displaystyle h_n > b_n $ and  $ \displaystyle\frac{b_n}{\sqrt{c}•} < h_n.$ So that one has:

\begin{equation}                                        
\omega_n  = h_n \sqrt{1- \frac{b_n^2}{h_n^2•}}\geq  h_n \sqrt{1-c} 
\end{equation}
\noindent

and hence 
\begin{equation}
\sum_{n=N_c}^\infty \frac{e^{-t(h_n-\omega_n)}}{\omega_n} \leq \frac{2 \ell^2\,\,\varepsilon }{\pi^2\sqrt{1-c}•} \,\, \xi (2)\,\, e^{\,\frac{1}{\varepsilon+\alpha(\ell/\pi)^2}}. 
\end{equation}

Besides, if $ \displaystyle \ell\geq 2\pi/a(1+\sqrt{1-\alpha \varepsilon)},  $    terms $\displaystyle \sum_{n=0}^{N_1-1} \frac{e^{-t(h_n-\omega_n)}}{\omega_n } $ have to  be considered.
 
 Since
  $ \displaystyle\sqrt{1-\alpha \varepsilon } = 1- \alpha \varepsilon /2 - (\alpha \varepsilon)^2 \,\,(1-\theta \alpha \varepsilon )^{-3/2} /8\,\,(0<\theta <1)   $ there exists a positive constant  $ B $ such that:

\begin{equation}
\sum_{n=0}^{N_1-1} \frac{e^{-t(h_n-\omega_n)}}{\omega_n}\leq  B  \, m \biggl(1+ e^{\,\frac{1}{\varepsilon+\alpha(\ell/\pi)^2}}\biggr) \varepsilon   
\end{equation}

being $ m  $ the minimum value of  $\displaystyle \frac{\omega_n•}{\varepsilon^2} $.

Otherwise, when $ N_1 <1 $ attention must be paid to $\displaystyle \sum_{n=0}^{N_2-1} \frac{e^{-h_n t}
 \sinh(\omega_nt)}{\omega_n}. $ If $  \displaystyle 1<q<\infty $ and $1/p+1/q=1 $   Holder inequality can be considered:

\begin{equation}
\sum_{n=0}^{N_2-1} \frac{e^{-h_n t}
 \sinh(\omega_nt)}{\omega_n}\leq \biggl(\sum_{n=0}^{N_2-1} |e^{-h_n t} \sinh(\omega_nt)|^p\biggr)^{1/p} \,\, \biggl(\sum_{n=0}^{N_2-1}  \left |\frac{1}{\omega_n}  \right |^q\biggr)^{1/q}.
\end{equation}

So, having
 
\[ \displaystyle \left |  \frac{1}{\omega_n} \right| ^q  \,\,\leq \frac{\varepsilon^q}{ \sqrt{1-\alpha \varepsilon}}
\] 
it is possible to find a positive constant  $D $  such that the following inequality holds:

\begin{equation}
\sum_{n=0}^{N_2-1} \frac{e^{-h_n t}
 \sinh(\omega_nt)}{\omega_n}\leq D\,\, \varepsilon^{\frac{q-1}{q}} \frac{1}{\sqrt{1-\alpha \varepsilon}}.
\end{equation}

In this way  (\ref{A1}) holds.

As for the x-differentiation of Fourier series like  
(\ref{A3}), attention must be given to convergence problems. Therefore, we
 consider x-derivatives of the operator
$(\varepsilon\partial_t+1)G$       instead of   $G $ and   $G_t$. Following \cite{mda} theorem  can be completely proven.

\hbox{} \hfill \rule{1.85mm}{2.82mm}

\end{document}